\documentclass[12pt]{article}
\usepackage{geometry} 
\usepackage{graphicx}
\usepackage[parfill]{parskip}
\usepackage{amssymb}
\usepackage{epstopdf}

\title{A simple formula for gravitational MHV amplitudes}
\author{Andrew Hodges\thanks{andrew.hodges@wadh.ox.ac.uk, http://www.twistordiagrams.org.uk. }\\{\footnotesize  {\it Mathematical Institute, University of Oxford, Oxford OX1 3LB, U.K.}}}

\date{9 April 2012}
\begin{document}
\maketitle
\begin{abstract}
A simple formula is given for the $n$-field tree-level MHV gravitational amplitude, based on soft limit factors. It expresses the full $S_n$ symmetry naturally, as a determinant of elements of a symmetric $(n\times n)$ matrix.\end{abstract}

\section{Introduction}

This note extends the material introduced in (Hodges 2011). That paper showed how BCFW recursion (Britto, Cachazo, Feng and Witten 2005) can be applied with N=7 super-symmetry to write down simplified expressions for all tree-level gravitational amplitudes. In particular, for MHV amplitudes this method yielded a recursion relation which avoids spurious double poles and gives a direct proof of the standard BGK expressions (Berends, Giele and Kuijf 1988), later  justified on quite different grounds by Mason and Skinner (2009). Further sections of the paper developed a calculus of phase factors in which shorter and suggestive expressions for 6- and 7-field MHV amplitudes were given.

 Now we pursue this program further by proving a new formula for the $n$-field MHV amplitude. This, the analogue of the Parke-Taylor formula in gauge theory, effects a great simplification. This new result also clarifies the momentum-twistor picture introduced in (Hodges 2011), by proving the existence of polynomials which express the content of the gravitational interaction.
\vspace{5mm}

\section{The phase factor and soft factor definitions}

In (Hodges 2011) the following useful definition of phase factors was given:
\begin{eqnarray}\psi^i_j &=& \frac{[ij]}{\langle ij\rangle} \,\,({\hbox{for  }} i \ne j)\nonumber\\
\psi^i_i &=& 0 \, .
\label{eq:phidefn}\end{eqnarray}
We now make a different definition, using the symbol $\phi$ to avoid confusion with the $\psi$. For $i\ne j$ the definition is just the same, but when $i=j$ we make a significant change:
\begin{eqnarray}\phi^i_j &=& \frac{[ij]}{\langle ij\rangle} \,\,({\hbox{for  }} i \ne j)\nonumber\\
\phi^i_i &=& - \sum_{j\ne i} \frac{[ij]\langle jx\rangle \langle jy\rangle}{\langle ij\rangle\langle ix\rangle \langle iy\rangle}
\, .\label{eq:phidefn}\end{eqnarray}
This new quantity is the (negative of the) {\em universal gravitational soft factor} associated with adding the $i$th field to the others, as defined in (Nguyen, Spradlin, Volovich and Wen, 2009).  Momentum conservation ensures that the definition is independent of the spinors $x, y$. Note that the spinorial weight of the $\phi^i_j$ is $(-2)$ on each index, this remaining true for $\phi^i_i$.
It is most important to note that $ \phi^i_i$ is only defined relative to a complete set of $n$ momenta summing to zero; it has an implicit dependence on the the other $(n-1)$ momenta. 

This negative sign is chosen so that we have a convenient form for the vital linear relation, from which everything flows: 
\begin{equation}\sum_{j=1}^n \phi^i_j\, \pi_j^{A'}\pi_j^{B'} = 0\, .\label{eq:conservation}\end{equation}
It is also convenient to define
\begin{equation}c_{ijk} = c^{ijk} = \{ \langle ij\rangle\langle jk\rangle\langle ki\rangle\}^{-1}\, ,\end{equation}
so that the $c_{ijk}$ are completely antisymmetric in their indices. 

As in (Hodges 2011), we shall use square brackets round indices to indicate anti-symmetrization without any $1/n!$ factor.

\section{The new formula}

Then the main result is that the reduced gravitational MHV amplitude $\bar{M}_n$ is given simply by:
\begin{equation}\bar{M}_n(12\ldots n) = (-1)^{n+1}\,\hbox{sgn}(\alpha\beta)\,c_{\alpha(1)\alpha(2)\alpha(3)}c^{\beta(1)\beta(2)\beta(3)}\phi_{[\alpha(4)}^{\beta(4) }\phi_{\alpha(5)}^{\beta(5)} \dots \phi^{\beta(n)}_{\alpha(n)]}\, ,\label{eq:main}\end{equation}
where $\alpha$ and $\beta$ are any permutations of $\{123\dots n\}$.

The $\pm 1$ factors for the signature of the permutations are obviously necessary. Otherwise, the overall sign is not of great importance, as the definition of the reduced amplitude is conventional. But the $(-1)^{n+1}$ ensures that $\bar{M}_3(123)$ is simply $c_{123}c^{123}$ and that for $n>3$ the formula is consistent with the definition of $\bar{M_n}$ given by the recursive relation in (Hodges 2011), as we shall soon show.

Equivalently, let $\Phi$ be the $n\times n$ symmetric matrix formed by the $\phi^i_j$, and $|\Phi|_{ijk}^{rst}$ be the $(n-3)\times (n-3)$ minor determinant obtained by striking out rows $i, j, k$ and columns $r,s,t.$ Then
\begin{equation}\bar{M_n}(12\ldots n) = (-1)^{n+1}  \sigma(ijk,rst)\,c^{ijk}c_{rst} |\Phi|_{ijk}^{rst}\, ,\label{eq:main2}\end{equation}
where $\sigma(ijk,rst) = \hbox{sgn}((ijk12\dots i\!\!\!/j\!\!\!/k\!\!\!/\dots n) \rightarrow (rst12\dots  r\!\!\!/s\!\!\!/t\!\!\!/\dots n)).$

We first establish that  formulas (\ref{eq:main}) and   (\ref{eq:main2}) are well-defined, i.e.\ that they are independent of the permutations, and so enjoy $S_n$ symmetry. We first show that
\begin{equation}c_{123} |\Phi|^{123}_{rst} = - c_{124}|\Phi|^{124}_{rst}\, \label{eq:main32}\end{equation}
 Note that $\{r,s,t\}$ may overlap with $\{1,2,3, 4\},$ without restriction.

To do this, it is useful to define $f_j^i=  \langle i1\rangle \langle i2\rangle\phi^i_j$. Then by (\ref{eq:conservation}),
\begin{equation}\sum_{i=1}^n f_j^i = 0\, .\label{eq:rowsum}\end{equation} That is, the rows in the complete $n\times n$ matrix $f_j^i$ all sum to zero.
The identity to be shown is equivalent to:
$$ |f|^{123}_{rst} = - |f|^{124}_{rst}\, ,$$
which is immediate from (\ref{eq:rowsum}) and the elementary properties of determinants. But now similarly $c_{ijk} |\Phi|^{ijk}_{rst} = - c_{ijm}|\Phi|^{ijm}_{rst}$, and then any permutation can be composed from such transpositions. This completes the proof, and presents the $S_n$ symmetry as a trivial consequence of (\ref{eq:conservation}).

The expression (\ref{eq:main}) is thus well-defined, completely symmetric, and also of the right spinorial weight. It remains to show that  it satisfies the recursion relation as derived in (Hodges 2011), at equation (59):
\begin{equation}
\bar{M}_n(123\ldots n-1,n) = \sum_{p=3}^{n-1} \frac{[pn]}{\langle p n\rangle}\frac{\langle1p\rangle\langle 2p\rangle}{\langle1n\rangle\langle 2n\rangle}\bar{M}_{n-1}(\hat{1}_{(p)} 23\ldots \hat{p}\ldots n-1)\, ,\label{eq:recursion}\end{equation}
where \begin{equation}\hat{1}_{(p)}] =  \frac{ (1+n)|p\rangle}{ \langle1p\rangle}, \quad\quad \hat{1}_{(p)}\rangle = 1\rangle, \quad\quad \hat{p}] = \frac{(p + n)|1\rangle}{ \langle p1\rangle}, \quad\quad \hat{p}\rangle =p\rangle \, ,\end{equation}
so that $\hat{1}_{(p)} + \hat{p} = 1 + p + n$.
The notation $\hat{1}_{(p)}$ is used to emphasise that the shifted momentum $\hat{1}$ is different in each of the $(n-3)$ terms, depending on $p$. 

We can exploit the freedom of representation offered by the new formula to choose a helpful representation of the $\bar{M}_{n-1}$ at each point of the recursion. The algebraic complexity arises mainly from the `shifted' momenta $\hat{1}_{(p)}$ and $\hat{p}$, so we craftily put these within the set of three which do not appear in the determinant. In fact we choose the triple $\{12p\}$ for both  the excluded rows and the excluded columns. 
We also note that $$ \frac{[pn]}{\langle p n\rangle} = \phi^n_p, \quad\quad \frac{\langle1p\rangle\langle 2p\rangle}{\langle1n\rangle\langle 2n\rangle}c_{12p} = c_{12n}\,,$$
so that the consistency of the recursion relation  (\ref{eq:recursion}) is equivalent to showing:
\begin{equation}\bar{M}_n =  (-1)^n \sum_{p=3}^{n-1}\phi^n_p c_{12n}c^{12p} |\hat{\Phi}|^{12p}_{12p}\, .\end{equation}
Here the hatted $\hat{\Phi}$ is an $(n-1)\times(n-1)$ matrix in which the objects $\hat{\phi}^i_j$ are defined with respect to the $(n-1)$ shifted momenta $\{\hat{1}_{(p)}, 2, 3\dots p-1, \hat{p}, p+1,\dots n-1\}$ summing to zero. Our choice of representation means that the only difference between $\hat{\phi}^i_j$ and ${\phi}^i_j$ is that  within the $p$th term, 
\begin{equation}\hat{\phi}_k^k = \phi_k^k + \phi_n^k\frac{\langle1n\rangle\langle pn\rangle}{\langle1k\rangle\langle pk\rangle}\, .\end{equation}
Again it is useful to write:
\begin{equation}f^i_j =  \langle1i\rangle \langle 2i\rangle \phi^i_j\, ,\quad \hat{f}^i_j =  \langle1i\rangle \langle 2i\rangle \hat{\phi}^i_j\,,\quad \hat{f}^k_k = f^k_k + f^k_n\frac{\langle pn\rangle\langle2k\rangle}{\langle 2n\rangle\langle pk\rangle}\, .\end{equation}
We define $F$ as the $(n-3)\times (n-3)$ matrix with entries $f^i_j$ for $3 \le i, j\le n-1$, with its minor determinants indicated in the same way as for $\Phi$, and $\hat{F}$ analogously. So in these terms, it is required to prove that:
\begin{equation}(-1)^n\prod_{k=3}^{n-1} \langle1k\rangle\langle2k\rangle \, \,\bar{M}_n = c_{12n}c^{12n} \sum_{p=3}^{n-1}f^n_p |\hat{F}|^{p}_{p}\, . \end{equation}
To do this, we expand the $f^n_p|\hat{F}|^{p}_{p}$, ordering the sum  by  the number of shift-correction factors used in each term of the expansion. The zeroth order term has no correction terms, and gives $f^n_p|F|^{p}_{p}.$ A typical first order term comes from taking just one correction, say for $f^q_q$ in one of the terms of the summation, say the $p$th, where $p\ne q$. This contributes
$$ f^n_pf^n_q |F|^{pq}_{pq}\frac{\langle pn\rangle\langle2q\rangle}{\langle 2n\rangle\langle pq\rangle}\, .$$
Summing over all $p$ and all $q$, and so symmetrising over $p$ and $q$, the Schouten identity gives  this simple expression for the sum of all first-order corrections:
$$\sum_{3\le p<q\le n-1}f^n_pf^n_q |F|^{pq}_{pq}\, .$$
A typical second order term comes from taking two corrections in the $p$th term of the summation, say from  $f^q_q$ and  $f^r_r$, thus contributing
$$ f^n_pf^n_q f^r_r|F|^{pqr}_{pqr}\frac{\langle pn\rangle\langle2q\rangle}{\langle 2n\rangle\langle pq\rangle}\frac{\langle pn\rangle\langle2r\rangle}{\langle 2n\rangle\langle pr\rangle}\, .$$
Adding in the contribution from  $f^p_p$ and  $f^r_r$ in the $q$th term of the summation, and $f^p_p$ and  $f^q_q$ in the $r$th term of the summation, it is clear that every disjoint set $\{p,q,r\}$ contributes
$$ f^n_pf^n_q f^r_r|F|^{pqr}_{pqr}\frac{\langle pn\rangle\langle2q\rangle}{\langle 2n\rangle\langle pq\rangle}\frac{\langle pn\rangle\langle2r\rangle}{\langle 2n\rangle\langle pr\rangle} + (p\leftrightarrow q) + (p \leftrightarrow r)\, ,$$
which is readily seen to be
$$ f^n_pf^n_q f^r_r|F|^{pqr}_{pqr}\, ,$$
and so contributing a total
$$\sum_{3\le p<q<r\le n-1}f^n_pf^n_qf^r_r |F|^{pqr}_{pqr}\, .$$
The same simplification occurs at every order,
so it  remains only to show that:
\begin{equation} (-1)^n\prod_{k=3}^{n-1} \langle1k\rangle\langle2k\rangle \,\,\bar{M}_n = c_{12n}c^{12n}\sum_{i=1}^{n-3}\sum_{3\le p_{1}<p_2\dots p_i \le n-1} f^n_{p_1}\dots f^n_{p_i} |F|^{p_1p_2\dots p_i }_{p_1p_2\dots p_i }\, . \label{eq:expansion}\end{equation}
But this too is simple. Consider the $(n-3)\times(n-3)$ matrix $H$ with entries
$$h^i_j= f^i_j + \delta^i_j f^n_j \hbox{ for } 3\le i, j \le n-1\, .$$
Thus $H$ has the same entries as $F$, but with the addition of $f^n_i$ to elements down the main diagonal. Each row of $H$ sums to zero, by (\ref{eq:rowsum}), so its determinant $|H|$ vanishes. On the other hand, we can also expand $|H|$ in terms of the number of $f^n_i$ factors. The zeroth order part is just $|F|$. The $i$th order part is 
$$\sum_{3\le p_{1}<p_2\dots p_i \le n-1} f^n_{p_1}\dots f^n_{p_i} |F|^{p_1p_2\dots p_i}_{p_1p_2\dots p_i}\, .$$
It follows that
$$0 = |H| = |F| + \sum_{i=1}^{n-3}\sum_{3\le p_{1}<p_2\dots p_i \le n-1} f^n_{p_1}\dots f^n_{p_i} |F|^{p_1p_2\dots p_i}_{p_1p_2\dots p_i}\, .$$
Thus the claim we are checking reduces to
$$(-1)^n\prod_{k=3}^{n-1} \langle1k\rangle\langle2k\rangle \, \bar{M}_n = - c_{12n} c^{12n} |F| \, ,$$
which indeed is true, being equivalent to
$$\bar{M}_n = (-1)^{n+1}c_{12n}c^{12n} |\Phi|^{12n}_{12n}\, .$$
This observation concludes the recursive proof of the new formula for $\bar{M}_n$.

It is striking that whilst determinants are naturally thought of as generating {\em anti-symmetry}, the minor determinants of the symmetric $\Phi$ matrix naturally yield a $S_n$-{\em symmetry} --- exactly as needed for a gravitational amplitude. This suggests scope for  generalization beyond MHV tree amplitudes.  

The new formula is much simpler than the BGK-Mason-Skinner expression, in a very concrete sense. Given numerical data for the spinors, it requires only $O(n^2)$ operations to find the $\phi^i_j$, after which the determinant of a symmetric matrix of order $(n-3)$ must be computed. This is easily achieved in (better than) $O(n^3)$ time.  In contrast, the BGK-Mason-Skinner formula requires summation over $(n-3)!$ terms and so grows exponentially. The new gravitational formula even compares well with gauge theory, where the simplicity of the Parke-Taylor formula emerges only after the separation into $(n-1)!/2$ colour-order sectors, all of which must be considered. For a general gluon interaction, therefore, the complexity is exponential in $n$. This formula might be seen as an indication of the emergent simplicity of gravitational scattering, notably advanced by Nima Arkani-Hamed, Freddy Cachazo and Jared Kaplan (2008). It can hardly be doubted that further enormous simplifications can be achieved.

\section {Illustrative examples}

The new formula includes all the expressions given in (Hodges 2011) in terms of the $\psi^i_j$ and extends them by changing to the $\phi^i_j$. Thus  we have 
\begin{eqnarray}\bar{M}_3(123) &=& \frac{1}{\langle12\rangle \langle23\rangle\langle31\rangle\,\, \langle12\rangle \langle23\rangle\langle31\rangle}\,,\nonumber \\
\bar{M}_4(1234) &=& \frac{\phi^1_4}{\langle12\rangle \langle23\rangle\langle31 \rangle\,\, \langle23\rangle\langle34\rangle\langle42\rangle}\,,\nonumber\\
\bar{M}_5(12345) &= &\frac{\phi^1_{[4}\phi^2_{5]}}{\langle12\rangle \langle23\rangle\langle31 \rangle\,\,\langle34\rangle\langle45\rangle\langle53\rangle}\,,\end{eqnarray}
but now we also have, for instance,
\begin{eqnarray}\
\bar{M}_4(1234) &=&- \frac{\phi^4_4}{\langle12\rangle \langle23\rangle\langle31 \rangle\,\, \langle12\rangle\langle23\rangle\langle31\rangle}\,,\nonumber\\
\bar{M}_5(12345) &= &\frac{\phi^4_{[4}\phi^5_{5]}}{\langle12\rangle \langle23\rangle\langle31 \rangle\,\,\langle12\rangle\langle23\rangle\langle31\rangle}\,.\end{eqnarray}

Likewise for $n=6$ we have the new expression found in (Hodges 2011):
\begin{equation}\bar{M}_6(123456) =  \frac{\phi^1_{[4}\phi^2_5\phi^3_{6]}}{\langle12\rangle \langle23\rangle\langle31\rangle\,\,\langle45\rangle\langle56\rangle\langle64\rangle}\, ,\end{equation}
but also:
\begin{eqnarray}\bar{M}_6(123456) &=& -\frac{\phi^4_{[4}\phi^5_5\phi^6_{6]}}{\langle12\rangle \langle23\rangle\langle31\rangle\,\,\langle12\rangle\langle23\rangle\langle31\rangle}\, \nonumber \\
&=& \frac{\phi^4_{[4}\phi^5_5\phi^3_{6]}}{\langle12\rangle \langle26\rangle\langle61\rangle\,\,\langle12\rangle\langle23\rangle\langle31\rangle}\nonumber\\
&=& -\frac{\phi^4_{[4}\phi^2_5\phi^3_{6]}}{\langle15\rangle \langle56\rangle\langle61\rangle\,\,\langle12\rangle\langle23\rangle\langle31\rangle}\, .\end{eqnarray}

For $n=7$, expression (77) in (Hodges 2011) was offered as the shortest identifiable formula:
\begin{eqnarray}\bar{M}_7(1234567) &=& \frac{\psi^6_{[3}\psi^{1}_{4}\psi^2_{5]}\,\,\psi^7_{6}}{\langle12\rangle\langle27\rangle\langle71\rangle \,\,\langle34\rangle \langle45\rangle\langle53\rangle} +  \frac{\psi^3_{[4}\psi^{1}_{5}\psi^2_{6]}\,\,\psi^7_{3}}{\langle12\rangle\langle27\rangle\langle71\rangle \,\,\langle64\rangle \langle45\rangle\langle56\rangle}\nonumber\\&+&  \frac{\psi^4_{[3}\psi^{1}_5\psi^2_{6]}\,\,\psi^7_{4}}{\langle12\rangle\langle27\rangle\langle71\rangle \,\,\langle36\rangle \langle65\rangle\langle53\rangle}+  \frac{\psi^5_{[3}\psi^{1}_4\psi^2_{6]}\,\,\psi^7_{5}}{\langle12\rangle\langle27\rangle\langle71\rangle \,\,\langle34\rangle \langle46\rangle\langle63\rangle}\nonumber\\ &+&  \frac{\psi^7_{[3}\psi^{1}_4\psi^2_{5}\psi^6_{7]}}{\langle12\rangle\langle26\rangle\langle61\rangle \,\,\langle34\rangle \langle45\rangle\langle53\rangle}\, .\label{eq:7graviton}\end{eqnarray}
By making copious use of the 6-point identity noted in equation (65) of (Hodges 2011), this can 
be rewritten with a common denominator:
\begin{eqnarray}\bar{M}_7(1234567) &=& \frac{\psi^6_{[3}\psi^{1}_{4}\psi^2_{5]}\,\,\psi^7_{6}}{\langle12\rangle\langle26\rangle\langle61\rangle \,\,\langle34\rangle \langle45\rangle\langle53\rangle} \frac{\langle16\rangle\langle26\rangle}{\langle17\rangle\langle27\rangle}\nonumber\\&+&  \frac{\psi^6_{[3}\psi^{1}_{4}\psi^2_{5]}\,\,\psi^7_{3}}{\langle12\rangle\langle26\rangle\langle61\rangle \,\,\langle34\rangle \langle45\rangle\langle53\rangle}\frac{\langle13\rangle\langle23\rangle}{\langle17\rangle\langle27\rangle}\nonumber\\&+&  \frac{\psi^6_{[3}\psi^{1}_4\psi^2_{5]}\,\,\psi^7_{4}}{\langle12\rangle\langle26\rangle\langle61\rangle \,\,\langle34\rangle \langle45\rangle\langle53\rangle}\frac{\langle14\rangle\langle24\rangle}{\langle17\rangle\langle27\rangle}\nonumber\\&+&  \frac{\psi^6_{[3}\psi^{1}_4\psi^2_{5]}\,\,\psi^7_{5}}{\langle12\rangle\langle26\rangle\langle61\rangle \,\,\langle34\rangle \langle45\rangle\langle53\rangle}\frac{\langle15\rangle\langle25\rangle}{\langle17\rangle\langle27\rangle}\nonumber\\ &+&  \frac{\psi^7_{[3}\psi^{1}_4\psi^2_{5}\psi^6_{7]}}{\langle12\rangle\langle26\rangle\langle61\rangle \,\,\langle34\rangle \langle45\rangle\langle53\rangle}\, .\label{eq:7graviton2}\end{eqnarray}
But the first four terms are now naturally gathered together into the universal soft factor $\phi^7_7$, so giving
\begin{eqnarray}\bar{M}_7(1234567) &=& \frac{\psi^6_{[3}\psi^{1}_{4}\psi^2_{5]}\,\,(- \phi^7_{7})}{\langle12\rangle\langle26\rangle\langle61\rangle \,\,\langle34\rangle \langle45\rangle\langle53\rangle} +  \frac{\psi^7_{[3}\psi^{1}_4\psi^2_{5}\psi^6_{7]}}{\langle12\rangle\langle26\rangle\langle61\rangle \,\,\langle34\rangle \langle45\rangle\langle53\rangle}\nonumber\\
&=&-\frac{\phi^1_{[3}\phi^{2}_4\phi^6_{5}\phi^7_{7]}}{\langle12\rangle\langle26\rangle\langle61\rangle \,\,\langle34\rangle \langle45\rangle\langle53\rangle}\, .\label{eq:7point}\end{eqnarray}
in agreement with the new general formula.
It was actually this observation that suggested the definition of $\phi^i_i$, and hence the extension to $n>7$.

   \newpage
\section{The momentum-twistor numerator}

It was conjectured in (Hodges 2011), at equation (97), that  a natural polynomial arises when we express the $n$-point gravitational MHV amplitude in terms of the momentum-twistor space introduced in (Hodges 2009). That is, we define
$N_n(12\ldots n)$ by
\begin{equation}\tilde{M}_n(123\ldots n) = \frac{N_n(123\ldots n)}{\prod_i \langle i, i+1\rangle \,\, \prod_{i < j} \langle ij\rangle}\, ,\end{equation}
and the conjecture is that $N_n$ is a polynomial (rather than a rational function).

The proof of this conjecture now follows immediately from cunning choice of the representation
\begin{equation}\bar{M}_n(12\ldots n) =(-1)^n \frac{\phi^1_{[2}\phi^{3}_{4}\psi^5_{6}\phi^7_{7}\phi^{8}_{8}\phi^9_{9}\ldots \phi^n_{n]}}{\langle13\rangle \langle35\rangle\langle51\rangle\,\,\langle24\rangle\langle46\rangle\langle62\rangle}\, .\end{equation}
Translating this into momentum twistors as defined by the ordering $(123\ldots n)$, it is obvious that just two denominator factors of $\langle 12\rangle$ will occur in each term of the expansion. These are safely absorbed in $\prod_i \langle i, i+1\rangle \,\, \prod_{i < j} \langle ij\rangle$, and the numerator function is therefore not singular in  $\langle 12\rangle$. By cyclicity, it cannot be singular in any other $\langle i, i+1\rangle$. A similar argument applies to all the other $\langle ij\rangle$ factors, and so it must be a polynomial.

For $n=5$ it was shown in (Hodges 2011) that the polynomial is the area of a pentagon defined by the 5 points in $\mathbb{C}^2$ defined by $\langle1234\rangle\pi_5^{A'}$ etc. The general polynomial is of degree $(n-3)$ in the twistors and degree $(n-3)(n-4)/2$ in $I$. So it is a $(n-3)$-degree polynomial in the $n!/(n-4)!4!$ objects like $\langle1234\rangle \pi_5^{A'}\ldots \pi_n^{N'}$.
It remains to be seen how this can be characterized geometrically.

\section{Acknowledgements}
These investigations were stimulated by my May-June 2010 visit to the Institute for Advanced Study, thanks to an invitation from Nima Arkani-Hamed, and I am most grateful  for generous support from that institution. A second visit in May  2011 allowed  further highly stimulating discussions. The interest   and encouragement of Marcus Spradlin and Anastasia Volovich has  been notable. However, it is Nima Arkani-Hamed who has constantly urged the importance of the soft factors. The new formula is just a sum over different ways of composing such factors, and it is indebted to his insight.

\section{References}

N. Arkani-Hamed, F. Cachazo and J. Kaplan, What is the simplest quantum field theory?, arXiv:0808.1446 (2008)

F. A. Berends, W. T. Giele and H. Kuijf, Phys. Lett. B {\bf 211}, 91 (1988). 

R. Britto, F. Cachazo, B. Feng and E. Witten, Direct proof of tree-level recursion relation in Yang-Mills theory, Phys. Rev. Lett. {\bf 94}, 181602, arXiv:hep-th/0501052 (2005)
 
A. Hodges, Eliminating spurious poles from gauge-theoretic amplitudes, \newline  arXiv:0905.1473 (2009)

A. Hodges, New expressions for  gravitational scattering amplitudes, \newline arXiv:118.2227v2 (2011)

L. J. Mason and D.  Skinner, Gravity, twistors and the MHV formalism, \newline arXiv:0808.3907v2 (2009)

D. Nguyen, M. Spradlin, A. Volovich and C. Wen, The tree formula for MHV graviton amplitudes, arXiv:0907.2276v2 (2009)

\end{document}